\documentclass{KapProc}
\usepackage{cite,amsmath}
\renewcommand{\theequation}{\arabic{section}.\arabic{equation}}

\renewcommand{\(}{\left( }
\renewcommand{\)}{\right) }
\renewcommand{\[}{\left[ }
\renewcommand{\]}{\right] }

\newcommand{\dlim}{\displaystyle\lim}
\newcommand{\Exp}{\operatorname{Exp}}
\newcommand{\lc}{\lowercase}
\newcommand{\ti}{\textit}
\newcommand{\Tr}{\operatorname{Tr}}

\newtheorem{conjecture}{Conjecture}

\begin{document}
\booktitle[Symbolic Computation]
{Symbolic Computation, Number Theory, Special Functions, Physics and
Combinatorics}
\articletitle[$\lc{q}$-Random Matrix Ensembles]{$\lc{q}$-Random Matrix Ensembles}
\setcounter{page}{1}
\chaptitlerunninghead{$\lc{q}$-Random Matrix Ensembles}

\author{K. A. Muttalib}
\affil{Department of Physics, University of Florida, Gainesville, Florida 32611-8440}
\author{Y. Chen}
\affil{Department of Mathematics, Imperial College, 
180 Queen's Gate, London SW7 2BZ, United Kingdom}
\author{M. E. H. Ismail}
\affil{Department of Mathematics, University of South Florida, Tampa, Florida 33620-5700}

\begin{abstract}
Theory of Random Matrix Ensembles have proven to be a useful tool in the
study of the statistical distribution of 
energy or transmission levels of a wide variety of physical systems. We give an
overview of certain $q$-generalizations of 
the Random Matrix Ensembles, which were first introduced in connection with
the statistical description of 
disordered quantum conductors.
\end{abstract}

\section{Introduction: Random Matrix Ensembles}

With a few notable exceptions, the interaction between the community 
of mathematicians who work in special functions, in particular, those
that are in the area of $q$-series and basic Hypergeometric functions and
the physics community has so far
been minimal. In this review article, we will describe some developments in
one area in physics, namely the Theory
of Random Matrix Ensembles, where a $q$-generalization has been proposed to
be relevant for multifractal states near a critical regime. The term 
multifractal states refers to the novel fractal characteristics of the 
quantum mechanical wave functions of
a disordered electronic conductor. It is well known in the physics literature
that depending on the strength of the disorder, a quantum conductor may
undergo a phase-transition from a metallic state which describes a good
conductor to those of an insulating state where the (static) conductivity  
in a finite sample is very small. In a region near the transition such
multifractal wave functions are observed at least in numerical experiments.

In addition to giving an overview of the results
obtained so far as well as providing some
details that have not been published before, the hope of the authors 
is to draw the attention of mathematicians to some mathematically posed 
unsolved problems that are important from the physics point-of-view. We expect
that this will also provide some additional insight into some of the 
asymptotic behavior of the $q$-polynomials
that may not be obvious from their conventional definitions. More precisely, 
we suspect that there exists certain ``universal scaling limits,'' 
to be explained later, for at least a class of
$q$-polynomials in certain asymptotic regime that have not been fully explored
by those mathematicians who are interested in $q$-special functions. 
The work discussed here were
done over last several years in many collaborations with Carsten Blecken, 
John Klauder, Vladimir Kravtsov and Nikos Nicopoulos.

Originally, Random Matrix Ensembles (RMEs) were proposed by E. Wig-ner in the
early sixties to describe the statistical 
properties of the eigenvalues and eigenfunctions of complex many-body
quantum systems in which the Hamiltonians of such systems are only 
defined in a probabilistic setting \cite{Mehta}. Over the
past fifteen years, they proved to be a useful tool in the studies of
equilibrium and transport properties of
disordered quantum systems, classically chaotic systems with a few degrees
of freedom, two-dimensional gravity,
conformal field theory and chiral phase transitions in quantum
chromodynamics \cite{Guhr}. The reason that such 
a wide variety of physical systems (with sizes ranging over more than nine
orders of magnitude between a nucleus
and a micron size  disordered quantum conductor) can be described by theory
of RMEs, which has no adjustable
parameter in it, is that there exists a special \ti{double scaling limit} in
which the correlations between
appropriately scaled eigenvalues in all these systems are universal,
independent of any parameter that describes the particular system.

In this section, we will briefly introduce the standard \ti{Gaussian} RME,
without any attempt to show how the 
expressions are derived. Detailed derivations can be found in \cite{Mehta}.
The idea is to point out the reason for
the existence of the \ti{zero-parameter} universality under certain double
scaling in Gaussian RME. We will then 
introduce the so-called ``Coulomb gas'' picture to just point out how the
zero parameter universality had been
``proven'' to be valid \ti{beyond} Gaussian RME. These will also serve the
purpose of introducing notations and
terminologies. In the following section we will show how this ``proof''
breaks down when the parameter $q$ ($0<q<1$) is introduced. This 
then points towards to a new \ti{one-parameter} universality. We will then
review various  known results as well as
open problems for some of the $q$-ensembles. Except for the section on
circular ensembles, we will consider 
only unrestricted Hermitian matrices relevant for physical systems with
broken time-reversal symmetry (which for example described a charged
system in the presence of a magnetic field). 
The corresponding ensemble is known as the unitary ensemble. 
We would like to point out here that although the Gaussian orthogonal and 
symplectic ensembles are by now very well understood largely due to the 
work of Tracy and Widom, the corresponding
$q$-generalizations have not been studied to date.

\subsection{Gaussian RME}

Consider the set of all $N\times N$ Hermitian matrices $M$ randomly chosen
with the following probability measure
\begin{equation}
\label{eq1}
P^{G}_N(M)dM=A_N\exp[-\Tr(M^2)]dM,
\end{equation}
where $A_N$ is a constant, $\Tr$ is the matrix trace and $dM$ the Haar 
measure. By going over to the eigenvalues-eigenvectors representation, it
can be shown that the joint probability distribution of the
eigenvalues $X=(x_i, i=1,2,\dots N)$ of the matrices can then be written
(up to $V_N$, the volume of the unitary group) in the form \cite{Mehta}
\begin{equation}
\label{eq2}
P^{G}_N(X)dx_1 \dots dx_N=
C^G_N \prod_{1\leq i<j \leq N}^N (x_i - x_j)^2 \prod_{i=1}^N e^{-x^2_i}.
\end{equation}
Here the factor $\prod_{i<j}(x_i-x_j)^2$ arises from the Jacobian of  
a change of variable, see \cite{Deift}. 

The level correlations can be determined exactly by recognizing that the
distribution can be written as a product of
Vandermonde determinants of a set of (monic) orthogonal Hermite polynomials. 
The orthonormal Hermite polynomials 
satisfy the orthogonality
relation
\begin{equation}
\label{eq3}
\int\limits_{-\infty}^{\infty} e^{-x^2}H_n(x)H_m(x)dx=
\delta_{mn}. 
\end{equation}
The main quantity of interest is the two-level kernel $K_N (x,y)$ from which
all correlation functions can be 
obtained. For example the $n$-level correlation function is given by
\cite{Mehta}
\begin{equation} 
\label{eq4}
\begin{split}
R_n(x_1,x_2,\dots x_n)
&=\frac{N!}{(N-n)!}\int\limits_{-\infty}^{\infty}\dots\int\limits_{-\infty}^{\infty} P_N(x_1,x_2,\dots
x_N)dx_{n+1}\dots dx_N\\
&=\det [K_N(x_i,x_j)]_{i,j=1,2\dots n}
\end{split}
\end{equation}
For the above Gaussian distribution the kernel has the explicit form
\begin{equation}
\label{eq5}
\begin{split}
K^G_N(x,y) &=e^{-(x^2+y^2)/2}\sum_{n=0}^{N-1} H_n(x)H_n(y) \\
&=e^{-(x^2+y^2)/2}\sqrt{N/2}\,
\frac{[H_N(x)H_{N-1}(y)-H_N(y)H_{N-1}(x)]}{(x-y)}
\end{split}
\end{equation}
where we have used the Christoffel-Darboux formula for the last equality. In
particular, the density of levels can be
obtained simply as $K_N(x,x)$. 
The Plancherel-Rotach asymptotics for the Hermite polynomials 
treat the large $N$ behavior if 
$x = \sqrt{2N+1}\cos \phi$, or $x = \sqrt{2N+1} \cosh \psi$, where $\phi$ 
and $\omega$ are fixed and 
$$
0 < \epsilon \le \phi\le \pi -\epsilon,  
\quad 0 < \omega \le \psi \le \omega,
$$
see \cite{Szego}. 
In the large $N$ limit, with $x$ and $N$ as above, the density of levels 
\begin{equation}
\label{eq6}
\sigma_N(x)=K^G_N(x,x)\sim\frac{1}{\pi}\sqrt{2 N - x^2}, 
\end{equation}
and zero otherwise.

We will now consider the double scaling limit, given by $N\to\infty$, $x$, 
$y\to 0$ such that the products
$\frac{\sqrt{2N}}{\pi}\,x=\zeta$, $\frac{\sqrt{2N}}{\pi}\,y=\eta$ remain finite. 
This particular $N$
dependence of the product is chosen so that
the density $\overline{\sigma}(\zeta)\equiv\sigma_N(x)dx/d\zeta=1$ in the new
scaled variable. We will ignore here
rapidly falling density at the edges $\(\pm\sqrt{2N}\)$ where it becomes 
zero and have focused our discussion to the middle of the density, i.e., $x$
near $0$ where $\sigma_N(0)=\frac{\sqrt{2N}}{\pi}$. In these new 
scaled variables, i.e., $\xi$ and $\eta$ and using the known asymptotics
of Hermite polynomials, the two-level ``limit kernel'' becomes
\begin{equation}
\label{eq7}
\overline{K}_G(\zeta,\eta)= \frac{\sin[\pi
(\zeta-\eta]}{\pi(\zeta-\eta)}; \;\;\;
\overline{\sigma}_G(\zeta)=\overline{K}_G(\zeta,\zeta)=1.
\end{equation}
This kernel, which is universal because it does not depend on $N$, gives
rise to the famous Wigner-Dyson-Mehta correlations of the eigenvalues \cite{Mehta}. 
Note that if we choose the ensemble from a more general Gaussian 
distribution
\begin{equation}
\label{eq8}
P_N(M)=A_N\Exp[-\Tr(aM^2+bM+c)],
\end{equation}
the parameters $a$, $b$, $c$ can always be removed by a simple completing of
squares. Therefore by an analogous rescaling the 
universality (parameter independence) of the two-level limit kernel is 
again reproduced. The details of this treatment are in 
Percy Deift's excellent monograph \cite{Deift}. 

\subsection{Beyond Gaussian RME: The Coulomb Gas Picture}

We now choose a non-Gaussian distribution of the ensemble defined by
\begin{equation}
\label{eq9}
P^V_N(M)=A^V_N\Exp[-Tr(V(M))]
\end{equation}
where $V(x)$ is a suitably increasing function of $x$, so that
the multiple integral,
\begin{equation}
\int\limits_{-\infty}^{\infty}\dots \int\limits_{-\infty}^{\infty}P^V_N(X) \; dx_1 
\dots dx_N, \nonumber
\end{equation}
exists for any fixed $N$, where the joint
probability distribution of the eigenvalues is
\begin{equation}
\label{eq10}
P^V_N(X)=C^V_N \prod_{i < j}^N (x_i - x_j)^2 \prod_{i=1}^N e^{-V(x_i)}.
\end{equation}
By going through the same 
procedure as described above we can write the two-level
kernel in terms of a set of polynomials that are orthogonal with respect to
the weight function $w(x)=\exp(-V(x))$
rather than $e^{-x^2}$ which previously led to the Hermite polynomials.
However, for an arbitrary $V(x)$, one would not expect the corresponding 
orthogonal polynomials to have explicit representations, 
or simple properties, and therefore this is usually not
a very useful description. Dyson
introduced a Coulomb gas picture \cite{Dyson} by writing
\begin{equation}
\label{eq11}
P^V_N(X)= e^{\beta H},
\end{equation}
and identifying
\begin{equation}
\label{eq12}
H=-\sum_{i <j}^N\ln|x_i-x_j|+\frac{1}{\beta}\sum_{i}^N V(x_i)
\end{equation}
with the Hamiltonian of a ``gas'' of $N$ classical charges on a
line interacting with each other with
logarithmic Coulomb repulsion and held together by a 
\ti{confining potential} $V(x)$, at a ``temperature''
$1/\beta$. For our unitary ensembles, $\beta=2$. In the large $N$ limit,
Dyson assumed that a continuum description is valid. 
(We will see later that this is where the $q$-ensembles differ in a 
subtle way). This is a plausible approximation if the density of the
charges increases with $N$ 
so that the spacing between the charges
decreases with $N$, and the $N\to\infty$
limit allows one to use a continuum approximation. In this case, the
two-level correlation function (related to the
two-level kernel by eq. \eqref{eq4}) can be written as \cite{Beenakker94}
\begin{equation}
\label{eq13}
R_2(x,y)= \frac1\beta\,\frac{\delta\sigma(x)}{\delta V(y)}.
\end{equation}
The variational derivative can be evaluated in the coulomb gas picture by
minimizing the energy (subject to the 
positivity of the density), which leads to the integral equation 
\begin{equation} 
\label{eq14}
\delta V(x)+\int\limits_{J}\delta\sigma (y)\ln|x-y| \, dy =\text{constant},
\end{equation}
where $J\subset {\bf R}$ is the support of the density $\sigma(x).$ In
general, the determination of $J$ is a highly non-trivial problem
and remain unsolved. 
The inverse of the equation then shows that the functional derivative of
\eqref{eq13}, and therefore \ti{all level
correlations are independent of the choice of the form of the potential}, so
that the non-Gaussian ensemble has 
the same universality as the Gaussian ensemble provided that the
assumption that $J$ is a single interval is valid. This is the reason why many
physical systems, which may be
described by different constraints and hence different form for the
distribution of the ensemble, still follow the
universal zero-parameter description of the RME at the double scaling limit.
Note that the proof depends only on 
the existence of the continuum limit where the electrostatic coulomb gas
picture holds.

\subsection{Beyond Gaussian RME: The Orthogonal Polynomials}

We now reconsider the non-Gaussian case from the point-of-view of the
orthogonal polynomials. As mentioned
before, for any confining potential $V(x)$, we can define the set of
polynomials $\phi_n(x)$ orthogonal with respect to the
weight function $w(x)=e^{-V(x)}$,
\begin{equation}
\label{eq15}
\int\limits_{-\infty}^{\infty} e^{-V(x)}\phi_n(x)\phi_m(x)dx
=
\delta_{mn},
\end{equation}
such that the Christoffel-Darboux kernel becomes
\begin{equation}
\label{eq16}
K^V_N(x,y)= e^{-(V(x)+V(y))/2}\sum_{n=0}^{N-1} \phi_n(x)\phi_n(y).
\end{equation}
The point now is that although we don't know the polynomials for arbitrary
$V(x)$, the universality in the coulomb 
gas picture implies that polynomials orthogonal with respect to {\it any} weight
function  must have the scaling property in
the asymptotic region qualitatively similar to the  Hermite polynomials,
namely 
\begin{equation}
\label{eq17}
\begin{split}
\dlim_{N\to\infty} (-1)^N N^{1/4}H_{2N}(x)& =\pi^{-1/2}\cos \pi\zeta;  \\ 
\dlim_{N\to\infty} (-1)^N N^{1/4}H_{2N+1}(x)& =\pi^{-1/2}\sin\pi\zeta, 
\end{split}
\end{equation} 
such that the double scaling that removes the $N$-dependence from the
density (to make it unity in the scaled 
variable) also removes the $N$-dependence from the two-level kernel. In
other words, the universality of RMEs for
some confining potentials is related to some ``universal'' asymptotic
scaling properties of the orthogonal
polynomials. 

\setcounter{equation}{0}

\section{$\lc{q}$-Orthogonal Polynomials: Failure of the Zero-Parameter Universality}

Clearly, various random matrix models can be defined either by choosing a
confinement potential, or equivalently 
a set of orthogonal polynomials. It is therefore obvious to ask if the
$q$-polynomials share the same scaling
characteristics as the classical polynomials. Historically, physicists
working with random matrix ensembles did not
know about $q$-polynomials, so this question was never asked. It was a
particular weight function arising in the
problem of disordered conductors \cite{Muttalib87,Stone,Slevin} that
eventually led the first and second named authors to seek  
help from Prof. George Andrews who suggested that Mourad Ismail be brought 
on board, starting a very fruitful collaboration. 

In the next section, we will consider some specific examples 
of the $q$-ensembles. Here we mention the important general
features of $q$-polynomials that are relevant for the breakdown of the
double scaling universality for the
$q$-ensembles. 

For classical orthogonal polynomials, the density $\sigma_N(x=0)$ scales
with some power of $N$ (e.g.,
$\sigma_N(x=0)\propto\sqrt{N}$ for Hermite polynomials). It has two
consequences. First, the spacing between
nearest two levels scales as a power of $1/N$, so that the continuum limit
can be achieved by taking the
$N\to\infty$ limit. Second, the fact that
$\dlim_{N\to\infty}\sigma_N(0)\to\infty$ guarantees that if the moments of the
weight function  
$$
\mu_j=\int\limits_{-\infty}^{\infty}x^jw(x)dx, 
$$
are given then $w$ is uniquely determined.
For those $q$-polynomials that correspond to the indeterminate moments
problem, the density $\sigma_N(0)$ is
independent of $N$. 
The fact that in this case 
$\dlim_{N\to\infty}\sigma_N(0)$ is finite implies that the corresponding
moment problem is indeterminate\cite{Akhiezer}.  Clearly we expect the 
asymptotics of these polynomials to have different scaling properties.
We have observed however that all the cases of $q$-orthogonal polynomials 
when the measure (weight) of 
orthogonality is not unique one can always find a weight function $w(x)$ 
so that $(\ln w(x))/\ln^2 x$ has a limit as $|x| \to \infty$. This probably 
contains a new universality law. This distinguishes $q$-orthogonal 
polynomials, that is polynomials whose recursion coefficients in the 
orthonormal form grow exponentially, from other cases of indeterminacy 
with recursion coefficients growing at a polynomial rate. 

It turns out that the weight functions for some $q$-polynomials behave as
$e^{- c\ln^2(x)}$ for large
$x\to\infty$ for some positive constant $c$. An indication of this can 
be seen from the growth of the
the recurrence coefficients as in the conjecture formulated at the 
end of this section.  If the recurrence coefficients
of a symmetric set 
of orthogonal polynomials 
$a_N$ grow as ${\rm O}(q^{-a N})$, for $a > 0$ then the 
$V(x)={\rm O}((\ln x)^2)$, see  
\cite{Ismail} for an example. 
As the moment problem is indeterminate, there are different
choices of the weight function possible for
the same recurrence relation. However, $V(x)\propto\ln^2(x)$ for large $x$
is a characteristic property for all
choices. Comparing with the Coulomb gas picture, this means that the
confinement potential in this case is ``weak,''
comparable to the repulsive interaction $\ln|x-y|$, as opposed to a ``strong''
power law confinement in case of
classical polynomials which always dominate over the logarithmic repulsion.
Note here that in order for the moment problem to be determinate $V(x)$ 
should grow at least as fast as $|x|$ as $x$ tends to infinity.
This difference is crucial. In fact, if 
we choose a weak confinement potential of the form $V(x)=\ln^2(1+ax)$ for
which no explicit (or closed) form of the orthogonal polynomial is 
known, the level correlations obtained by numerical methods turn out to be
the same as those obtained analytically
from exact solutions of the $q$-models \cite{Canali}. Thus the soft confinement,
$N$-independent spacing of the levels and 
indeterminate moments are all related to each other, and are generic
characteristics of what we will call the
``$q$-Random Matrix Ensembles.'' We will see later that these ensembles are
related to the physics of 
multifractality near the critical regime of a phase transition where the
wave function of a particle in a quantum
conductor changes from an extended state to an exponentially localized state
when randomness is increased.

The following conjecture is based on observations we made over the years. 
$$
$$

\begin{conjecture}
Let $p(x)$ be symmetric orthonormal polynomials satisfying 
$$
xp_n(x) = a_{n+1} p_{n+1}(x) + a_n p_{n-1}(x)
$$
and assume that $a_n q^{an}$ has a limit as $n\to \infty$, for some $a >0$. If 
$w$ is a weight function for the $p_n$'s then   
$$
\ln w(x) \approx -c \ln^2 |x|, \quad\text{for some } c > 0,
$$
as $x \to \infty$ or $x \to -\infty$, but not necessarily for both.
\end{conjecture}

In fact we even believe that the conjecture holds even for discrete measures in 
the sense that the $w(x_n)$,  the mass at $x_n$ satisfies the asymptotic 
relation $\ln w(x_n) \approx -c \ln^2 |x_n|$. There are ample examples to illustrate
 this behavior, see \cite{Ismail}. 

\subsection{The $\lc{q}$-Hermite RME}

Following \cite{Muttalib93}, we will define the $q$-Hermite random matrix
model by the $q$-Hermite polynomials
orthogonal with respect to the Askey weight function $w_A$ \cite{Askey}
\begin{equation}
\label{eq18}
\begin{split}
&\int\limits_{-\infty}^{\infty} h_n(\sinh u;q) h_m(\sinh u;q)w_A(u;q)\, du \\
& \qquad =q^{-n(n+1)/2}(q,q)_n(q;q)_{\infty}\, \ln(1/q)\, \delta_{nm}
\end{split}
\end{equation}
where
\begin{equation}
\label{eq19}
w_A(u;q)=\frac1{\(-q e^{-2u},-q e^{2u};q\)_{\infty}}; \;\;\; 0<q < 1.
\end{equation}
Here $-\infty < u <\infty$, and
\begin{equation}
\label{eq20}
(a; q)_{n}=\prod_{i=1}^{n}\(1-aq^{i-1}\); \quad
(a_1,a_2,\dots a_n;q)_m = \prod_{i=1}^n(a_i,q)_{m},
\end{equation}
where $m =0, 1, \dots$ or $\infty$. 
Ismail and Masson \cite{Ismail} referred to the 
polynomials $h_n$  as the $q^{-1}$-Hermite polynomials.
Note first of all that the polynomials in \eqref{eq18} are in the variable
$\sinh u$, which means that the
confinement potential has to be compared in the variable $x=\sinh u$. To
check the asymptotic behavior of the
potential, we use the Jacobi triple product formula \cite{Gasper}
\begin{equation}
\label{eq21}
\(q, -z \sqrt{q},  -\sqrt{q}/z,q\)_{\infty}
=\sum_{-\infty}^{\infty}q^{n^2/2}z^n.
\end{equation}
Choose $z= q^{1/2}e^{2u}$, giving
\begin{equation}
\label{eq22}
\frac{1+e^{-2u}}{w_A(u;q)}= 
\frac1{(q,q)_{\infty}}\sum_{-\infty}^{\infty} q^{n(n+1)/2}
e^{2 n}.
\end{equation}
Note that the right hand side of \eqref{eq22} is a constant multiple of a theta function. 
Also bear in mind that as a function of $x$ the weight function should be 
$w_A(u;q)/\cosh u$, and $x = \sinh u$. So
applying the Jacobi imaginary transformation, we find after going back to
the $x$ variable, 
\begin{equation}
\label{eq23}
V_A(x;q)={\rm O}((\sinh^{-1}x)^2/\gamma),
\end{equation}
which is ${\rm O}(\ln x)^2$. Here $\gamma=\ln(1/q).$

 From the asymptotic formulas derived by Ismail and Masson \cite{Ismail}, 
the Christoffel-Darboux
kernel is found in the limit $N\to\infty$ to be 
\begin{equation}
\label{eq24}
\begin{split}
K_A(u,v;q) &=\frac{\sqrt{w(u)w(v)}}{[(q;q)_{\infty}]^2\ln(1/q)}\\
& \qquad \times  
\(-qe^{u+v};-qe^{-u-v};qe^{u-v};-qe^{-u+v};q\)_{\infty}.
\end{split}
\end{equation}
Note that the density, in this $N\to\infty$
limit, is
\begin{equation}
\label{eq25}
\sigma_A(u;q)=K_A(u,u;q)=1/\ln(1/q).
\end{equation}
We note here that the Christoffel-Darboux kernel has a limit as 
$N\to\infty$ for fixed $u$ and $v$ and therefore the density also has a limit. 
Compare the above with the ordinary
Hermite case
\begin{equation}
\label{eq26}
\sigma_N(u=0,q =1)= \frac{\sqrt{2N}}{\pi}.
\end{equation}
Note that for any $q<1$, the $N\to\infty$ limit of the kernel exists. 
This is a discontinuous change, and
does not allow us to study the important question of how the zero-parameter
universality breaks down in a physical system. 

It is tempting to conjecture that the physically appropriate
generalization of the density from the $q=1$
case should involve
\begin{equation}
\label{eq27}
N_q\equiv \frac{1-q^N}{1-q}=\frac{1-e^{-\gamma N}}{1-e^{\gamma}};
\;\;\; q\equiv e^{-\gamma}
\end{equation}
replacing $N$, which is equal to $N$ for $\gamma N\ll 1$, but becomes
independent of $N$ for $\gamma N\gg 1$. This means that we 
should explore the asymptotic behavior of $q$-polynomials in the double
scaling limit where $\gamma\to 0$,
$N\to\infty$, but the product $\gamma N$ remains finite. Although they have
been used in certain special cases
\cite{Blecken98}, systematic studies of such double scaling limits of
$q$-polynomials have not been done. The kind of limits just described;
$q$ becoming $N$ dependent would require analysis involving varying weight.
This should be of interest to mathematicians who work in the
asymptotics of orthogonal polynomials.

The kernel \eqref{eq24} is independent of $N$, but depends on $q$. The
question then becomes, can the
$q$-dependence be scaled out by changing to variables such that the density
becomes unity, thereby restoring the
zero-parameter universality? The answer was given in \cite{Muttalib93}. Here
we provide some detailed steps of the
calculation (and supply the $q$-dependent constant factors). We first
express the infinite products in terms of
Jacobi theta functions $\theta_i$ \cite{Gasper},
\begin{equation}
\begin{split}
\(qe^{u-v};-qe^{-(u-v)};q\)_{\infty} &=\frac{q^{-1/8}}{2(q;q)_{\infty}}
\,\frac{\theta_1(i(u-v)/2;\sqrt{q})}{i\sinh((u-v)/2)} \\
& \quad  \times \(-qe^{u+v};-qe^{-(u+v)};q\)_{\infty} \\
 &=\frac1{2q^{1/8}(q;q)_{\infty}}\,
\frac{\theta_2(i(u+v)/2;\sqrt{q})}{i\cosh((u+v)/2)}.
\end{split}
\end{equation}
The kernel can be expressed as
\begin{equation}
\label{eq30}
K_A(u,v;q)=\frac{\Omega(u,v)
\Theta_2(u,v;\sqrt{q})}{2[(q;q)_{\infty}]^3\ln(1/q)q^{1/8}}\,
\frac{\theta_1(i(u-v)/2;\sqrt{q})}{i\sinh((u-v)/2)}
\end{equation}
where
\begin{equation}
\label{eq31}
\Omega(u,v)=\frac{\sqrt{\cosh u \cosh v}}{\cosh(u+v)/2}; \;\;\;
\Theta_2(u,v;\sqrt{q})=\frac{\theta_2(i(u+v)/2;\sqrt{q})}
{\sqrt{\theta_2(iu;\sqrt{q})\theta_2(i v;\sqrt{q})}}.
\end{equation}
We now use Jacobi's imaginary transformations for the theta functions
\cite{Whittaker},
\begin{equation}
\label{eq32}
\begin{aligned}
\theta_2\(i u;\sqrt{q}\)
&=\sqrt{\frac{2\pi}{\gamma}}\,e^{2u^2/\gamma}\theta_4\(\frac{2\pi
u}{\gamma};p\); \\
\theta_1\(iu;\sqrt{q}\) &=i\sqrt{\frac{2\pi}{\gamma}}\,e^{2u^2/\gamma}
\theta_1\(\frac{2\pi u}{\gamma},p\),
\end{aligned}
\end{equation}
where $p=e^{-2\pi^2/\gamma}$. The kernel now becomes
\begin{equation}
\label{eq33}
K_A(u,v;q)= a(q)\Omega(u,v)\,\frac{\theta_4(\pi(u+v)/\gamma;p)}
{\sqrt{\theta_4(2\pi u/\gamma;p)\theta_4(2\pi v/\gamma;p)}}\,
\frac{\theta_1(\pi(u-v)/\gamma;p)}{\sinh((u-v)/2)},
\end{equation}
where
\begin{equation}
\label{eq34}
a(q)=\sqrt{\frac{2\pi}{\gamma}}\,
\frac1{2\gamma}\,\frac{q^{-1/8}}{[(q;q)_{\infty}]^3 }.
\end{equation}
Defining new variables
\begin{equation}
\label{eq35}
\zeta=u/\gamma;\quad \eta=v/\gamma
\end{equation}
we finally have the appropriately scaled two-level kernel
\begin{equation}
\label{eq36}
\begin{split}
\overline{K}_A(\zeta,\eta;q) &=
a(q)\Omega(\gamma\zeta,\gamma\eta)\Theta_4(\zeta,\eta)
\,\gamma\frac{\theta_1(\pi(\zeta-\eta);p)}{\sinh((\zeta-\eta)\gamma/2)}; \\
\overline{K}_A(\zeta,\zeta;q)&=1,
\end{split}
\end{equation}
where
\begin{equation}
\label{eq37}
\Theta_4(\zeta,\eta) =\frac{\theta_4(\pi(\zeta+\eta);p)}
{\sqrt{\theta_4(2\pi \zeta;p)\theta_4(2\pi \eta;p)}}
\end{equation}
Apart from the prefactor $a(q)$, this is the limit kernel quoted in
\cite{Muttalib93}. Note that the kernel remains a 
function of $\gamma=\ln(1/q)$, so that the distribution is no longer
parameter independent. However, the
dependence on the levels as well as $q$ are quite complicated, and various
limiting cases turn out to be quite
interesting. In the rest of the section we will consider some of these
limits.

One corollary of the above expressions for the kernel before and after the
imaginary Jacobi transformations is that
we can equate the density $K_A(u,u)$ before and after the transformations to
obtain the relation
\begin{equation}
\label{eq38}
[(q;q)_{\infty}]^3=\(\frac{2\pi}{\gamma}\)^{3/2}q^{-1/8}p^{1/4}\[\(p^2;p^2\)
_{\infty}\]^3.
\end{equation}
Using this, and the product expansion of the Jacobi theta function
\begin{equation}
\label{eq39}
\theta_1(u;p)=2(p^2;p^2)_{\infty}p^{1/4}\sin u
\prod_1^{\infty}\[1-2p^{2n}\cos 2u +p^{4n}\],
\end{equation}
we can now consider some characteristics of the limit kernel.

For $\zeta=\eta+\epsilon$, we get
\begin{equation}
\label{eq40}
\overline{K}_A(\epsilon)\approx
\frac{\gamma}{2\pi}\,\frac{\sin[\pi\epsilon]}
{\sinh[\gamma\epsilon/2]}
\prod_{n=1}^{\infty}\[1+\frac{4p^{2n}}{(1-p^{2n})^2}\sin^2[\pi\epsilon]\]+
{\rm o}(1)\end{equation}
This expression depends only on the difference $\zeta-\eta$. We will see
that the lack of this ``translational invariance'' for the general 
expression \eqref{eq36} has important consequences. For $p\ll 1$, i.e., 
$\gamma/2\pi^2\ll 1$, the second term in the product can be neglected, and
we get the simple hyperbolic kernel
\begin{equation}
\label{eq41}
\overline{K}_A(\zeta-\eta; \gamma)=
\frac{\gamma}{2\pi}\,\frac{\sin[\pi(\zeta-\eta)]}
{\sinh (\gamma (\zeta-\eta)/2)}+{\rm o}(1); \;\;\;  \gamma/2\pi^2 \ll 1.
\end{equation}
The above limit kernel for the $q$-Hermite model for $\gamma/2\pi^2 \ll 1$
has been shown to be equivalent 
\cite{Kravtsov} to a banded random matrix model that 
describes multifractal wave functions at the critical regime 
for a quasi one dimensional disordered quantum conductor with weak
multifractality \cite{Mirlin}. Comparing the
two, \ti{the parameter $\gamma=\ln(1/q)$ can be explicitly identified with
the fractal dimensionality at the critical
regime} \cite{Kravtsov}. In contrast, the universal kernel resulting from
the ordinary Hermite polynomial describes 
chaotic systems with extended wave functions, i.e., those wave functions that 
extends to the edge of the sample. Thus the
$q$-deformation is in some way related to the physics of
multifractality, although the nature of the relation is not at all clear.
One major problem in understanding the
connection is that we do not know how to study the crossover from $q<1$ to
$q\to 1^{-}$. 
For a fixed $q<1$, the large $N$ limit kernel is independent 
of $N$. This prevents us from exploring in details how the double scaling
fails as one approaches the limit  
$q\to 1^{-}$ when the $q$-Hermite polynomials reduce to the ordinary Hermite
polynomials and the restoration of the zero-parameter universality. As 
pointed out before, an interesting direction is to study  
\ti{the double scaling regime $\gamma\to 0$, $N\to\infty$, 
keeping $\gamma N$ finite}, and then consider the limit kernel.
Properties of $q$-polynomials in such special double scaling limits
have not been explored yet. 

The situation is completely different in the case where $\eta$ is near 
$-\zeta$. The physical importance of this ``off-diagonal'' correlation 
was emphasized in \cite{Canali}. Consider $\eta=-\zeta$. Again for 
$p\ll 1$, the $\theta$-functions simplify, and we obtain
\begin{equation}
\label{eq42}
\overline{K}_A(\zeta,-\zeta)=\frac{\gamma}{2\pi}\coth(\gamma\zeta)
\sin(2\pi\zeta)+{\rm o}(1).
\end{equation}
Since $\coth\gamma\zeta\to 1$ for large $\zeta$, the kernel have long range
oscillations of frequency $2\pi$
whose amplitude grows linearly with $\gamma$ (as long as $\gamma\ll 2\pi^2$). 
The case for larger $\gamma$
should still have the oscillations, although the prefactor will change. In
other words, the kernel exhibits long range
correlations in the $\eta=-\zeta$ direction. Another physically interesting
limiting case is the kernel in the limit
$q\to 0$, i.e., $\gamma\to\infty$. The kernel was heuristically shown to be of
the form \cite{Bogomolny,Kravtsov}
\begin{equation}
\label{eq43}
\overline{K}_A(\zeta,\eta)\approx\begin{cases} 
1, & \text{if }\zeta\eta < 0; \\ 
(-1)^{[2\zeta]}, & \text{if }\zeta\eta > 0;
\end{cases}
\end{equation}
in squares of unit size along the two diagonals $\zeta=\eta$ and $\zeta
=-\eta$, where $[2\zeta]$ is the integer 
part of $2\zeta$. 
For all other values of $\zeta$ and $\eta$,
$K(\zeta,\eta)=0$. This is a surprising result. It is not
clear if this result is valid only for the $q$-Hermite polynomials, or is it
a generic feature of $q$-polynomials in the
$q\to 0$ limit. In particular, the asymptotic properties of the
$q$-polynomials for large $N$ in the scaling limit
$N\to\infty$, $\gamma\to\infty$, keeping the \ti{ratio} finite would be
interesting from physical point-of-view.

\subsection{An Alternative $\lc{q}$-Hermite Random Matrix Model}

Since the moment problem associated with the $q$-Hermite polynomials is
indeterminate, it is instructive to consider a
different $q$-Hermite model to check if the above results are sensitive to
the particular choice of the weight function. One such model is to choose
\begin{equation}
\label{eq44}
\int\limits_{-\infty}^{\infty} h_n(x;q)
h_m(x;q)w_{qH}(x;q)dx =\sqrt{\frac{\pi\gamma}
{2}}\,\frac{(q,q)_n}{q^{(n+1/2)^2/2}}\delta_{nm},
\end{equation}
where $w_{qH}$ is the weight function
onsidered in  \cite{Atakishiyev}
\begin{equation}
\label{eq45}
w_{qH}(x;q)=e^{-2(\sinh^{-1}x)^2/\gamma}.
\end{equation}
This model of $q$-RME was first considered in \cite{Muttalib96}. Here we
provide some details.

Note that the asymptotic behavior of the confinement potential in this model
is the same as for the Askey weight. 
Equation \eqref{eq44} can be rewritten in the variable $\zeta=\sinh^{-1}x/\gamma$
\begin{equation}
\label{eq46}
\begin{gathered}
\int\limits_{-\infty}^{\infty}h_n(\sinh \gamma\zeta;q) h_m(\sinh
\gamma\zeta;q)\cosh \gamma\zeta e^{-2\gamma\zeta^2}d\zeta \\
=\sqrt{\frac{\pi}{2\gamma}}\,\frac{(q,q)_n}{q^{(n+1/2)^2/2}}\; 
\delta_{nm},
\end{gathered}
\end{equation}
to take advantage of the Ismail-Masson result for the kernel, which becomes
\begin{equation}
\label{eq47}
\begin{split}
& \overline{K}_{qH}(\zeta,\eta;q) \\
& \quad =\sqrt{\frac{2\gamma}{\pi}}\,\frac{q^{1/8}}
{(q,q)_{\infty}}\sqrt{\cosh\gamma\zeta
\cosh\gamma\eta}\,e^{-\gamma(\zeta^2+\eta^2)} \\
&\qquad\times(-qe^{\gamma(\zeta+\eta)};-qe^{-\gamma(\zeta+\eta)}; 
qe^{\gamma(\zeta-\eta)};qe^{-\gamma(\zeta-\eta)}q)_{\infty}.
\end{split}
\end{equation}
In terms of Jacobi theta functions, the kernel has the form, after taking
the $N\to\infty$ limit,
\begin{equation}
\label{eq48}
\begin{split}
\overline{K}_{qH}(\zeta,\eta;q)
&=\sqrt{\frac{2\gamma}{\pi}}\,\frac1{2iq^{1/8}(q,q)_{\infty}}\\
& \quad \times  
\frac{\sqrt{\cosh\gamma\zeta\cosh\gamma\eta}}
{\cosh\gamma(\zeta+\eta)/2\sinh\gamma(\zeta-\eta)/2} \\
&\quad\times\theta_2[i\gamma(\zeta+\eta)/2;\sqrt{q}]
\theta_1[i\gamma(\zeta-\eta)/2;\sqrt{q}].
\end{split}
\end{equation}

Using the Jacobi imaginary transformation as before, we obtain
\begin{equation}
\label{eq49}
\overline{K}_{qH}(\zeta,\eta;q)=
c(q)\Omega(\gamma\zeta,\gamma\eta)\theta_4(\pi(\zeta+\eta))
\frac{\theta_1[\pi(\zeta-\eta);p]}{\sinh[(\zeta-\eta)\gamma/2]},
\end{equation}
where
\begin{equation}
\label{eq50}
c(q)=\sqrt{\frac{2\pi}{\gamma}}\,\frac{1}{2q^{1/8}[(q;q)_{\infty}]^3}
\end{equation}
and $\Omega$ has been defined before. Note that
$\overline{K}_{qH}(\zeta,\zeta)=\theta_4(2\pi\zeta)$ is not unity, so one 
needs to rescale the variables again. However, for $\gamma \ll 2\pi^2$, the
density is approximately unity, so we
can compare this kernel with the one from the  Askey weight, which is the
same in this regime. Moreover, it is also
the same for $\zeta=-\eta$. Heuristic arguments shows that the 
$q\to 0$ behavior to be similar for all potentials that have the same
large $x$ behavior \cite{Bogomolny}, so we expect it to be the same as well.
Note that in terms of the original
variable $x$, the confining potential had an oscillatory component in the
Askey weight while the density was
monotonic; in $q$-Hermite weight the potential is monotonic while the density is
oscillatory. Thus, the two models are very
different. It is only after the appropriate scaling that the kernel, and
therefore the physical properties, show similar
behavior. Note on the other hand that the $q$-Hermite weight is related to the Askey
weight as 
\begin{equation}
\label{eq51}
w_A(x;q)dx=\frac{2\gamma\sqrt{2(1-q)}(q,q)_{\infty}}
{\theta_4(2\pi\zeta;p)}\,w_{qH}(\zeta;q)d\zeta.
\end{equation}
Since the Jacobi theta function is periodic with respect to a shift
$\theta_4(z;q)=\theta_4(z+\pi;q)$, the two weights differ 
only by a periodic factor with respect to the shift $\zeta\rightarrow \zeta+m/2$ for
integer $m$. Whether the physical
correlations are always equivalent for all choices of the weight function
that correspond to the $q$-Hermite
polynomials, is an interesting open question.

\subsection{The $\lc{q}$-Laguerre RME}

We have already seen that the appropriately scaled physical correlations are
independent of the choice of the 
weight function within the $q$-Hermite random matrix model and appropriate
parameter regimes. In order to see 
if these are generic properties valid for other $q$-random matrices, we now
consider as an example the 
$q$-Laguerre RME, first considered in \cite{Chen1a,Chen1b}. It is defined by
the weight function \cite{Moak},
\begin{equation}
\label{eq52}
w_{qL}(x;q)=\frac1{(-(1-q)x;q)_{\infty}},
\end{equation}
with the orthogonality relation
\begin{equation}
\label{eq53}
\int\limits_{0}^{\infty} L_n(x;q) L_m(x;q)w_{qL}(x;q) dx
=\frac{\ln(1/q)}{(1-q)q^n}\,\delta_{nm},
\end{equation}
where $L_n(0;q)=1$. The orthogonality is on $x\ge 0$, so the
boundary at $x=0$ makes the model
qualitatively different from the $q$-Hermite model. Moreover, as shown in
\cite{Chen1a,Chen1b}, for small $x$ the confining potential is linear in $x$ 
as opposed to quadratic dependence for the $q$-Hermite potential. However, 
the large $x$ behavior is again $\ln^2 x$, so the breakdown of 
the zero-parameter universality should again be similar. In
particular, note that the density at $x=0$ can be obtained directly for any
$N$,
\begin{equation}
\label{eq54}
\sigma^{qL}_N(x=0;q)=(1-q^N)/\ln(1/q)=\frac{1-e^{-\gamma N}}{\gamma}.
\end{equation}
Compare this with $\sigma_N(x=0;q=1)=N$. As we suggested for the $q$-Hermite
model before, studying the
double scaling limit of the $q$-Laguerre polynomials $\gamma\to 0$,
$N\to\infty$ with the product
$\gamma N$ finite, would allow us to study the crossover regime from the
$N$-dependent density 
and the $N$-independent $q$-model.  
The $q$-Laguerre two-level kernel was first written down in
\cite{Blecken94}, and here we will provide some details. We use the
asymptotic properties of the $q$-Laguerre
polynomials \cite{Chen1a,Chen1b}
\begin{equation}
\label{eq55}
L_N(x;q)=
L_{\infty}(x;q)+\frac{q^{N+1}}{1-q}[L_{\infty}(x;q)-L_{\infty}(x/q;q)]
+{\rm O}(q^{N+2})
\end{equation}
where
\begin{equation}
\label{eq56}
L_{\infty}(x;q)=J^{(2)}_0\(2\sqrt{x(1-q)};q\).
\end{equation}
Here $J^{(2)}_0(z;q)$ is the $q$-Bessel function. Note that as before, the
$N\to\infty$ results are
independent of $N$. The kernel requires the combination
\begin{equation}
\label{eq57}
\begin{split}
L &\equiv L_{N+1}(y;q)L_{N}(x;q) -L_{N}(y;q)L_{N+1}(x;q) \\
&=q^{N+1}\[L_{\infty}\(y/q;q\)L_{\infty}(x;q)
-L_{\infty}(y;q)L_{\infty}\(x/q;q\)\]  \\
& \quad +\text{O}(q^{N+1}).
\end{split}
\end{equation}
Using the asymptotic expression for the $q$-Bessel function for $q\ll 1$,
\cite{Chen2}
\begin{equation}
\label{eq58}
\begin{split}
J^{(2)}_0\(2\sqrt{x(1-q)};q\) & \sim 
\sqrt{\frac{4\pi}{\gamma}}\cos\left\{(\pi/\gamma)\ln\sqrt{Qx}\right\}
\\
& \quad \times 
\exp\{[\ln^2\sqrt{Q x}- \pi^2/4]/\gamma\};\\
& Q  := 1-q,
\end{split}
\end{equation}
we get after some algebra
\begin{equation}
\label{eq59}
\begin{split}
L & =\frac{4\pi\sqrt{Q}}{\gamma}e^{-\pi^2/\gamma+\gamma/4}
e^{\[\ln^2 Q x+\ln^2 Q y\]} \\
&\; \; \times \left[\sqrt{x}\cos\left\{(\pi/2\gamma) \ln Q y\right\}
\sin\left\{(\pi/2\gamma)\ln Q x\right\} \right. \\
& \qquad \left. 
-\sqrt{y}\sin\left\{(\pi/2\gamma)\ln Q y\right\} 
\cos\left\{(\pi/2\gamma)\ln Q x\right \} \right]  +{\rm o}(1).
\end{split}
\end{equation}
For large $x$ such that $q \ll Q x$, we can use Jacobi's triple product
identity above to obtain
\begin{equation}
\label{eq60}
[w_{qL}(x;q)]^{-1}\sim\sum_{-\infty}^{\infty}q^{n^2/2}z^n;\quad 
z=\frac{Qx}{\sqrt{q}}.
\end{equation}
Approximating the sum by an integral we get
\begin{equation}
\label{eq61}
[w_{qL}(x;q)]=
\sqrt{\frac{\gamma}{2\pi}}e^{-\frac1{2\gamma}}[\ln Q x+\gamma/2]^2+{\rm o}(1).
\end{equation}
Using these expressions, we obtain the limit kernel
\begin{equation}
\label{eq62}
\begin{split}
K_{qL}(x,y) &=\frac{b(q)}{x-y}
\left[\(\frac{x}{y}\)^{1/4}\sin\[\frac{\pi}{2\gamma}\ln Q x\]
\cos\[\frac{\pi}{2\gamma}\ln Q y\]\right. \\
&\quad\left.-\(\frac{y}{x}\)^{1/4}
\sin\[\frac{\pi}{2\gamma}\ln Q y\]
\cos\[\frac{\pi}{2\gamma}\ln Q x\]\right]
\end{split}
\end{equation}
where
\begin{equation}
\label{eq63}
b(q)=\frac{\sqrt{8\pi}}{\gamma^{3/2}}\(\frac{p^2}{q}\)^{1/8},
\end{equation}
and $0<\gamma<2\pi$.
Introducing the variables $\zeta=\frac{\pi b(q)}{2\gamma}\ln Q x$ and
$\eta=\frac{\pi b(q)}{2\gamma}\ln Q y$, the kernel
becomes
\begin{equation}
\overline{K}_{qL}(\zeta,\eta)=\frac{\gamma}{\pi}\,
\frac{e^{\frac{\gamma(\zeta-\eta)}{2\pi b}}\sin(\zeta/b)\cos(\eta/b)
-e^{-\frac{\gamma(\zeta-\eta)}{2\pi b}}\sin(\eta/b)\cos(\zeta/b)}
{\sinh[(\zeta-\eta)\gamma/\pi b]}
\end{equation}
and $\overline{K}_{qL}(\zeta,\zeta)=1+(\gamma/2\pi)\sin(2\zeta/b)$. 
For $\gamma \ll 2\pi$,
the density is approximately unity and independent of $\zeta$
and we can compare the kernel with the $q$-Hermite model. 
For $\zeta\approx\eta$, $\overline{K}_{qL}(\zeta,\eta)$ has the same
form as $\overline{K}_A(\zeta,\eta)$ for the $q$-Hermite
model, and therefore the level correlations are similar. 

\subsection{$\lc{q}$-Circular RME}

For mathematical simplicity, Dyson introduced the Circular Ensemble where
the eigenvalues lie uniformly on the
complex unit circle \cite{Dyson2}. A possible $q$-generalization appropriate
for eigenvalues of scattering matrices
for disordered systems was introduced in \cite{Muttalib95a} defined by the
weight function
\begin{equation}
\label{eq65}
w_S(\theta;q)=\frac1{2\pi}\left|
\frac{(q^{1/2}e^{i\theta};q)_{\infty}}
{(aq^{1/2}e^{i\theta};q)_{\infty}}\right|^2;\quad 0<q<1,\quad a^2q <1.
\end{equation}
This defines the Szeg\H{o} polynomials generalized by Askey \cite{Szego},
orthogonal on the unit circle
\begin{equation}
\label{eq66}
\frac1{2\pi}\int\limits_{0}^{2\pi}\Phi_m(e^{i\theta})\,
\overline{\Phi_n(e^{i\theta})}\,w(\theta;q)d\theta=\delta_{mn}
\end{equation}
where the overline denotes complex conjugate. The two-level kernel is then
given by
\begin{equation}
\label{eq67}
K_S(\theta,\phi)
\sqrt{w_S(\theta;q)}\,\sqrt{w_S(\phi;q)}\sum_{n=0}^{N-1}
\overline{\Phi_n(e^{i\theta};q)}\,\Phi_n((e^{i\phi;q}).
\end{equation}
In the large $N$ limit, the density was obtained in \cite{Muttalib95a}
\begin{equation}
\label{eq68}
\sigma(\theta ;q)\approx \frac{N}{2\pi}
\left[1+\frac1{(1-q)N}\,\ln\, 
\left(\frac{1-2a\sqrt{q}\,\cos\theta+a^2q}{1-2a\sqrt{q}\,\cos\theta+q}\right)
\right].
\end{equation}
There are two important features of this result. First, the density has
finite $N$ correction to the uniform density
$N/2\pi$ of the $a=1$ Dyson circular ensemble only in the $q\to 1$ limit
such that the product $(1-q)N$ is kept
finite. Thus the physically interesting limit is again a double scaling
limit, although different from the previous 
ones. The second point is that although properly normalized, the density
becomes negative for sufficiently small
values of $\theta < \theta_c$, where $\theta_c$ depends on the parameter
$a$. (Physically, more disorder means
larger $\theta_c$). It was pointed out in \cite{Muttalib95a} that this
negative density arises as a result of the
opening up of a gap, which requires that this possibility be included from
the beginning by demanding 
orthogonality of the polynomials on the appropriate range
\begin{equation}
\label{eq69}
\frac{C}{2\pi}\int\limits_{\theta_c}^{2\pi-\theta_c}\Phi_m(e^{i\theta})\,
\overline{\Phi_n(e^{i\theta})}\,w(\theta;q)d\theta=\delta_{mn}.
\end{equation}
These are no longer the generalized Szeg\H{o} polynomials. The problem
suggests considering orthogonal
$q$-polynomials defined on \ti{parts} of a circle.

\subsection{Bi-orthogonal $\lc{q}$-RME}

The bi-orthogonal random matrix models were proposed in \cite{Muttalib95b},
where the joint probability
distribution is of the form
\begin{equation}
\label{eq70}
P_N(X)=C_N \prod_{m < n}^N\[r\(x_m\)-r\(x_n\)\]\[s\(x_m\)-s\(x_n\)\]
\prod_{n=1}^N w\(x_n\).
\end{equation}
Here $r(x)$ is a polynomial of degree $h$ and $s(x)$ is a polynomial of
degree $k$. The kernel is then given by 
\begin{equation}
\label{eq71}
K^B_N(x,y)=\sqrt{w(x)}\,\sqrt{w(y)}\sum_{n=0}^{N-1} Y_n(x)Z_n(y)
\end{equation}
where $Y$ and $Z$ are polynomials in $r(x)$ and $s(x)$ respectively,
satisfying the biorthogonality relation
\begin{equation}
\label{eq72}
\int\limits_I Y_n(x)Z_m(x)w(x)dx=g_n\delta_{mn}.
\end{equation}
Here $g_n$ is the normalization constant, and $I$ refers to the appropriate
range. It reduces to the ordinary 
ensembles for $r(x)=s(x)=x$. For the special case of $r(x)=x$, $s(x)=x^k$,
the biorthogonal limit kernel for the
ordinary Hermite and Laguerre ensembles were recently obtained by Borodin
\cite{Borodin} in terms of Wright's
generalized Bessel function. For the special case of $r(x)=x$, $s(x)=x^k$
and $w(x)=w_{qL}(x;q)$ (the $q$-Laguerre weight 
defined earlier), the polynomials are known as the $q$-Konhauser biorthogonal
polynomials \cite{AlSalam}. However, the
asymptotic form of the limit kernel for this model for large $N$ originally
proposed in \cite{Muttalib95b} has not
been obtained yet.

\setcounter{equation}{0}
\section{Other Physical Properties}

While the two-level limit kernel in principle determines all physical
correlation functions for unitary ensembles, 
evaluating a physically  interesting property is not always straightforward.
We mention two such properties
considered for $q$-RMEs so far.

\subsection{Fredholm Determinants for $\lc{q}$-RMEs}

An important quantity for studying the spacing distribution of adjacent
levels is the Fredholm determinant 
associated with the limit kernel. Tracy and Widom \cite{Tracy} considered
Fredholm determinants for general
random matrix models described by orthogonal polynomials. They showed that
under quite general conditions, 
one can obtain a system of partial differential equations for ensembles
defined by a weight function $w(x)$ if
$x^kw(x)$ is bounded for $k=0,1,\dots$ and $w(x)$ is continuously
differentiable within the interval. Nishigaki
\cite{Nishigaki}  has used this approach to obtain the diagonal resolvent of
the limit kernel \eqref{eq41}. This is
valid for $\gamma\ll 2\pi^2$, and has been shown to describe the spacing
distribution of disordered conductors
near the weakly multifractal metal to insulator transition region. (An
earlier attempt to describe the spacing
distribution was also limited to the weak multifractality regime
\cite{Muttalib96b}.) The corresponding result for 
the more general case valid for all $q$ has not been obtained yet. Numerical
results show a novel type of limiting
``triangular'' distribution in the limit $q\to 0$ \cite{Bogomolny}.

\subsection{Parametric Correlations}

In the presence of an external field $X$, the ``parametric'' level density
correlation function for $q$-Hermite model
was introduced in \cite{Blecken98} 
\begin{equation}
\label{eq73}
\begin{split}
S\(x,y;X,X'\) &=c w\(\sqrt{c}x\)w\(\sqrt{c}y\) \\
& \times \sum_{n=N}^{\infty}\sum_{m=0}^{N-1}
h_n\(\sinh\(\sqrt{c}x\);q\) h_n\(\sinh\(\sqrt{c}y\);q\) \\
& \times h_m\(\sinh\(\sqrt{c}x\);q\) h_m\(\sinh\(\sqrt{c}y\);q\)
e^{(m_q-n_q)\omega\(X-X'\)^2}
\end{split}
\end{equation}
where $c=\gamma\pi^2/2N_q$ and $\omega=c/\mu$ is related to a disorder
parameter. In the limit $X=0$, this
reduces to the ordinary two-level correlation function
\begin{equation}
\label{eq74}
S(x,y;X=0)= R_2(x,y)=\sigma (x)\sigma (y)-K^2(x,y).
\end{equation}
In the limit $q=1$ this reduces to the usual parametric density correlation
function \cite{Beenakker94}. The
expression \eqref{eq73} was considered in the special case $x=y=0$ only, in
the double scaling limit 
$\gamma\to 0$, $N\to\infty$, with the product $\gamma N$ finite. The general
case for all $x$, $y$ is not known. 
Other physically important quantities are the ``velocity correlator'' defined as
\begin{equation}
\label{eq75}
C\(E_1, E_2; X, X'\)=\int\limits_{}^{E_1}\!\int\limits_{}^{E_2} 
\frac{\partial^2}{\partial X\partial X'}S\(x,y; X,X'\)\,dxdy,
\end{equation}
and the `parametric number variance' defined as
\begin{equation}
\label{eq76}
U\(E_1, E_2; X\)=\int\limits_{}^{E_1}\!\int\limits_{}^{E_2} [S(x ,y; 0)-S(x, y; X)]\,dxdy
\end{equation}

Again only the limit $E_1=E_2=0$ has been considered in \cite{Blecken98,Blecken99}.

\section{Summary}

The two-level kernels (and other correlation functions) for $q$-ensembles
require the asymptotic behavior of 
various orthogonal $q$-polynomials and their weight functions for finite $N$
in various double scaling limits. Some
of these scaling properties seem to share common features in certain range
of parameters for apparently very
distinct $q$-polynomials.

\acknowledgments{
M. Ismail acknowledges partial support from the 
Liu Bie Ju Center of Mathematical Sciences at the City University of Hong Kong 
and NSF grants DMS-9970865 and INT-9713354.}

\normallatexbib


\begin{chapthebibliography}{10}

\bibitem{Akhiezer}
N.I. Akhiezer, \emph{The classical moment problem and some related questions in
  analysis}, Oliver and Boyd, Edinburgh, 1965 (English translation).

\bibitem{AlSalam}
W.A. Al-Salam and A.~Verma, {$q$-{K}onhauser polynomials}, \emph{Pacific J.
  Math.} \textbf{108} (1983), no.~1, 1--7.

\bibitem{Askey}
R.A. Askey, {Continuous $q$-{H}ermite polynomials when $q>1$}, in \emph{$q$-Series
  and partitions} (Minneapolis, MN, 1988) (D.~Stanton, ed.), Springer, New York,
  1989, pp.~151--158.

\bibitem{Atakishiyev}
N.M. Atakishiyev, A.~Frank, and K.B. Wolf, {A simple difference
  realization of the {H}eisenberg $q$-algebra}, \emph{J. Math. Phys.} \textbf{35}
  (1994), no.~7, 3253--3260.

\bibitem{Beenakker94}
C.W.J. Beenakker and B.~Rejaei, {Random matrix theory of parametric
  correlations in the spectra of disordered metals and chaotic billiards},
  \emph{Physica A} \textbf{203} (1994), 61--90.

\bibitem{Blecken99}
C.~Blecken, Y.~Chen, and K.~A. Muttalib, {Parametric number variance of
  disordered systems in the multifractal regime}, \emph{Waves in Random Media}
  \textbf{9} (1999), 83--90.

\bibitem{Blecken94}
C.~Blecken, Y.~Chen, and K.A. Muttalib, {Transitions in spectral
  statistics}, \emph{J. Phys. A: Math. Gen.} \textbf{27} (1994), no.~16, L563--L568.

\bibitem{Blecken98}
C.~Blecken and K.A. Muttalib, {Brownian motion model of a $q$-deformed
  random matrix ensemble}, \emph{J. Phys. A: Math. Gen.} \textbf{31} (1998), no.~9,
  2123--2132.

\bibitem{Bogomolny}
E.~Bogomolny, O.~Bohigas, and M.P. Pato, {On the distribution of
  eigenvalues of certain matrix ensembles}, \emph{Phys. Rev. E} (3) \textbf{55}
  (1997), no.~6, part A, 6707--6718.

\bibitem{Borodin}
A.~Borodin, {Biorthogonal ensembles}, \emph{Nucl. Phys. B} \textbf{536} (1998),
  704--732.

\bibitem{Canali}
C.M. Canali and V.E. Kravtsov, {Normalization sum-rule and spontaneous
  breaking of $u(n)$ invariance in random matrix ensembles}, \emph{Phys. Rev. E}
  \textbf{51} (1995), R5185--R5188.

\bibitem{Chen1a}
Y.~Chen, M.E.H. Ismail, and K.A. Muttalib, {A solvable random matrix model
  for disordered conductors}, \emph{J. Phys.: Condens. Matt.} \textbf{4} (1992),
  L417--L423.

\bibitem{Chen1b}
Y.~Chen, M.E.H. Ismail, and K.A. Muttalib,
{Metallic and insulating behavior in an exactly solvable random
  matrix model}, \emph{J. Phys.: Condens. Matt.} \textbf{5} (1993), 171--190.

\bibitem{Chen2}
Y.~Chen, M.E.H. Ismail, and K.A. Muttalib,
{Asymptotics of basic {B}essel functions and $q$-{L}aguerre
  polynomials}, \emph{J. Comp. Appl. Math.} \textbf{54} (1995), 263--273.

\bibitem{Deift}
P.A. Deift, \emph{Orthogonal polynomials and random matrices: a
  {R}iemann-{H}ilbert approach}, New York University, Courant Institute of
  Mathematical Sciences, New York, 1999.

\bibitem{Dyson}
F.J. Dyson, {Statistical theory of energy levels of complex systems {I}},
  \emph{J. Math. Phys.} \textbf{3} (1962), 140--156.

\bibitem{Dyson2}
F.J. Dyson, 
{Statistical theory of energy levels of complex systems {III}},
  \emph{J. Math. Phys.} \textbf{3} (1962), 166--175.

\bibitem{Gasper}
G.~Gasper and M.~Rahman, \emph{Basic hypergeometric series}, Cambridge
  University Press, Cambridge, 1990.

\bibitem{Guhr}
T.~Guhr, A.~Muller-Groeling, and H.~Weidenmuller, {Random matrix theories
  in quantum physics: common concepts}, \emph{Phys. Rep.} \textbf{299} (1998),
  no.~4-6, 189--425.

\bibitem{Ismail}
M.E.H. Ismail and D.R. Masson, {$q$-{H}ermite polynomials, biorthogonal
  rational functions and $q$-beta integrals}, \emph{Trans. Amer. Math. Soc.}
  \textbf{346} (1994), no.~1, 63--116.

\bibitem{Kravtsov}
V.E. Kravtsov and K.A. Muttalib, {New class of random matrix ensembles
  with multifractal eigenvectors}, \emph{Phys. Rev. Lett.}
 \textbf{79} (1997), 1913--1916.

\bibitem{Mehta}
M.L. Mehta, \emph{Random matrices}, second ed., Academic Press, Boston, MA,
  1991.

\bibitem{Mirlin}
A.D. {Mirlin, et al.}, {Transition from localized to extended eigenstates
  in the ensemble of power law random banded matrices}, \emph{Phys. Rev. E}
  \textbf{54} (1996), 3221--3230.

\bibitem{Moak}
D.S. Moak, {The $q$-analog of laguerre polynomials}, \emph{J. Math. Anal. Appl.}
  \textbf{81} (1981), 20--47.

\bibitem{Muttalib95b}
K.A. Muttalib, {Random matrix models with additional interactions}, \emph{J.
  Phys. A: Math. Gen.} \textbf{28} (1995), no.~5, L159--164.

\bibitem{Muttalib96}
K.A. Muttalib, 
{Transitions from {W}igner to {P}oisson distribution in a class
  of solvable models}, in \emph{Proceedings of the IV Wigner Symposium} (Guadalahara,
  1995) (River Edge, NJ) (N.M. Atakishiyev, T.H. Seligman, and K.B. Wolf,
  eds.), World Scientific Publishing, 1996, pp.~370--374.

\bibitem{Muttalib93}
K.A. Muttalib, Y.~Chen, M.E.H. Ismail, and V.N. Nicopoulos, {New family of
  unitary random matrices}, \emph{Phys. Rev. Lett.} \textbf{71} (1993), 471--475.

\bibitem{Muttalib95a}
K.A. Muttalib and M.E.H. Ismail, {Impact of localization on dyson's
  circular ensemble}, \emph{J. Phys. A: Math. Gen.} \textbf{28} (1995), no.~21,
  L541--548.

\bibitem{Muttalib96b}
K.A. Muttalib and J.R. Klauder, {Asymptotic level spacing distribution for
  a $q$-deformed random matrix ensemble}, \emph{J. Phys. A: Math. Gen.} \textbf{29}
  (1996), no.~16, 4853--4857.

\bibitem{Muttalib87}
K.A. Muttalib, J-L. Pichard, and A.D. Stone, {Random matrix theory and
  universal statistics for disordered quantum conductors}, \emph{Phys. Rev. Lett.}
  \textbf{59} (1987), 2475--2478.

\bibitem{Nishigaki}
S.M. Nishigaki, {Level spacings at the metal-insulator transition in the
  {A}nderson {H}amiltonian and multifractal random matrix ensembles}, \emph{Phys.
  Rev. E} \textbf{59} (1999), 2853--2862.

\bibitem{Slevin}
K.~Slevin, J.-L. Pichard, and K.A. Muttalib, {Maximum entropy-ansatz for
  transmission in quantum conductors: a quantitative study in $2$ and $3$
  dimensions}, \emph{J. Phys. France I} \textbf{3} (1993), 1387--1404.

\bibitem{Stone}
A.D. Stone, P.~Mello, K.A. Muttalib, and J.-L. Pichard, {Random matrix
  theory and maximum entropy models for disordered conductors}, in \emph{Mesoscopic
  Phenomena in Solids} (B.L. Altshuler, P.A. Lee, and R.A. Webb, eds.),
  North-Holland, Amsterdam, 1991, pp.~369--448.

\bibitem{Szego}
G.~Szeg{\H{o}}, \emph{Orthogonal polynomials}, fourth ed., American Math. Soc.,
  Providence, RI, 1975, American Mathematical Society, Colloquium Publications,
  Vol. XXIII.

\bibitem{Tracy}
C.A. Tracy and H.~Widom, {Fredholm determinants, differential equations
  and matrix models}, \emph{Comm. Math. Phys.} \textbf{163} (1994), no.~1, 33--72.

\bibitem{Whittaker}
E.T. Whittaker and G.N. Watson, \emph{A course of modern analysis}, Cambridge
  University Press, Cambridge, 1996. 
  Reprint of the fourth (1927) edition.

\end{chapthebibliography}

\end{document}